\documentclass[reprint,
aps,twocolumn]{revtex4}
\usepackage{amssymb}
\usepackage{amsmath}
\usepackage{graphicx}
\usepackage{bm}
\usepackage{psfrag}
\newcommand{\beq}{\begin{equation}}
\newcommand{\eeq}{\end{equation}}

\newcommand{\bee}{\begin{eqnarray}}
\newcommand{\eee}{\end{eqnarray}}
\newcommand{\ba}{\begin{array}}
\newcommand{\ea}{\end{array}}
\newcommand{\bc}{\begin{center}}
\newcommand{\ec}{\end{center}}
\newcommand{\bi}{\begin{itemize}}
\newcommand{\ei}{\end{itemize}}

\usepackage{color}
\usepackage{subfig}

\begin{document}

\title{
Dynamics of flexible fibers in shear flow
}

\author{Agnieszka M. S\l owicka} 
\author{Eligiusz Wajnryb}
\author{Maria L. Ekiel-Je\.zewska}
\email{mekiel@ippt.pan.pl}
\thanks{Corresponding author}
\affiliation{Institute of Fundamental Technological Research, Polish Academy of Sciences, Pawi\'nskiego 5b, 02-106 Warsaw, Poland}

\date{\today}

\begin{abstract}
Dynamics of flexible non-Brownian fibers in shear flow at low-Reynolds-number are analyzed numerically for a wide range of the ratios $A$ of the fiber bending force to the viscous drag force. Initially, the fibers are aligned with the flow, and later they move in the plane perpendicular to the flow vorticity. A surprisingly rich spectrum of different modes is observed  when the value of $A$ is systematically changed, with sharp transitions between coiled and straightening out modes, 
period-doubling bifurcations from periodic to migrating solutions, irregular dynamics and chaos. 
\end{abstract}

\maketitle


In the literature, there has been for decades a lot of interest in analyzing the dynamics of flexible non-Brownian and Brownian fibers in shear flow at low-Reynolds-number: experimentally, theoretically and numerically \cite{Forgas1,Forgas2,Yamamoto1993,Skjetne,Joung,Uesaka2007,slowicka_dynamics_2012}.
Recently, migration and rheology related to Brownian motion 
in shear and Poiseuille flows have been extensively investigated \cite{Jen,Ladd,Winkler2006,Winkler2010,Reddig,Yeomans}. The growing interest in this field has been motivated by its importance for applications for polymers, proteins, DNA or other biological systems. 

On the other hand, the coil-stretch transition of flexible polymers discussed by de Gennes \cite{deGennes} was followed by numerous fundamental theoretical and experimental studies of buckling instabilities of non-Brownian fibers in 
different ambient fluid flows
\cite{BeckerShelley,YoungShelley,Lindner,Kantsler,Harasim,slowicka_lateral_2013}. The results demonstrate a complexity of the dynamics of flexible fibers, and its sensitivity not only to the fiber bending rigidity, but also to many other different factors, such as e. g. curvature of the flow or presence of walls~\cite{slowicka_lateral_2013}, very large length of fibers leading to topological changes of their shapes~\cite{Doyle,Stone} or a spectrum of initial conditions leading to various modes of the three-dimensional motion \cite{Skjetne}. 

Therefore, in this work we focus on a  simple system: a single non-Brownian deformable fiber in an ambient shear flow of a fluid with viscosity $\eta$. The goal is to investigate the dependence of the  dynamics on the fiber  flexibility. We try to minimize or eliminate other sources of a complex behavior. Therefore, the flow is unbounded (there are no walls), 
the fiber is moving freely, and its aspect ratio is not large.   Initially, the fiber is straight, aligned with the flow, and there is no constraint forces acting on any of its segments, what leads to the fiber motion entirely 
within the plane $y\!=\!0$ perpendicular to vorticity of the ambient shear flow with velocity ${\bf v}_0= \dot{\gamma} z\hat{\bf x}$.

A fiber is modeled as a chain of 
$N$  identical solid spherical beads of diameter $d$ 
\cite{GaugerStark}. 
The length, time, force and velocity units are chosen respectively as 
\bee
d,\hspace{0.6cm}\tau =1/ \dot{\gamma},\hspace{0.6cm}{ f}_0= \pi \eta d^2\dot{\gamma},\hspace{0.6cm}{ \nu}_0=d/\tau.
\label{eq:force_unit}
\eee

The centers of the consecutive beads are linked by springs with the equilibrium length $l_0d$  
so small that the consecutive beads almost touch each other. The spring constant 
$kf_0/d$ is so large 
that the fiber's length practically does not change.

At the equilibrium, the fiber is straight. 
At a deformed configuration, there appear a bending force exerted on each bead, proportional to the (dimensional) bending parameter $Af_0d^2$ \cite{GaugerStark,BeckerShelley,slowicka_dynamics_2012,slowicka_lateral_2013}. Our goal is to analyze how the dynamics of fibers depends on their dimensionless bending stiffness $A$.

We assume that the Reynolds number is much smaller than unity, and 
the fluid velocity 
and pressure 
satisfy the stationary Stokes equations \cite{Kim}.
Velocities of the beads at given positions are 
evaluated 
with the use of the {\sc Hydromultipole} numerical {\sc fortran} code, based on the multipole method of solving the Stokes equations, with the multipole expansion
 truncated at an order $L$ \cite{Cichocki:1999}. The time-dependent positions of the beads are determined by the adaptive fourth order Runge-Kutta method. 
In addition, 
the time-step is reduced by a factor of two, if potential overlaps are detected, what allows to avoid spurious overlaps \cite{motyl}.

We choose the following values of the parameters, 
\bee
N\!=\!10, \hspace{0.3cm}  l_0 \!=\! 1.01, \hspace{0.3cm}  k\!=\!2000,  \hspace{0.3cm} 0.3 \!\le\! A \!\le\! 50, \hspace{0.3cm}L\!=\!2.\;\;
\eee

While moving along the flow, the fiber center-of-mass oscillates across the flow, 
with or without a systematic lateral migration, depending on the bending stiffness $A$. The fiber 
tumbles, in a similar way as a rigid elongated body 
\cite{Jeffery}. While it turns, its shape evolves accordingly.

\begin{figure*}
\includegraphics[width=18cm]{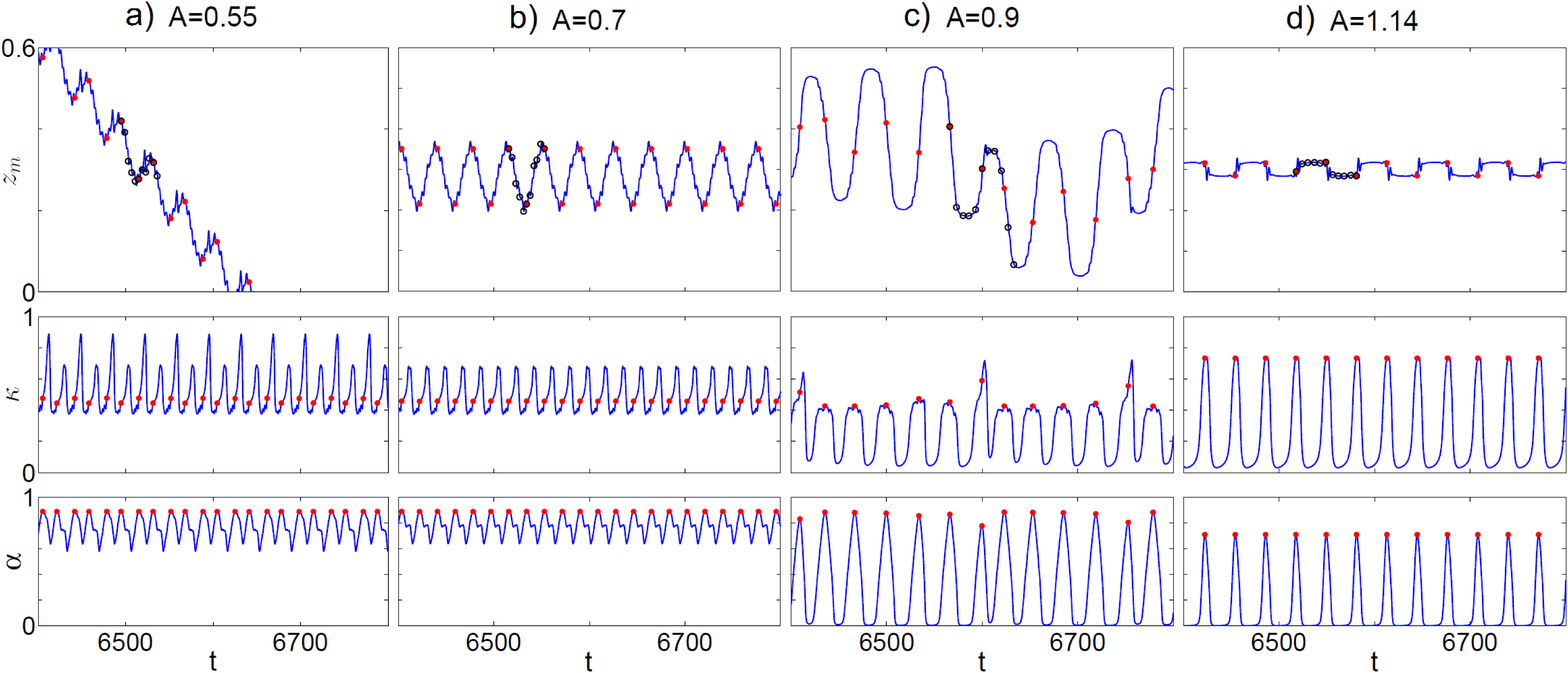}
\caption{Evolution of the fiber center-of-mass position 
$z_m$, curvature $\kappa$ and fractional compression $\alpha$ depends on the fiber 
stiffness $A$. Red dots indicate flipping times, and black circles - times of the corresponding snapshots in Fig.~\ref{fig2b}. }
\label{fig3a}
\end{figure*}

To determine the lateral motion, we evaluate $z_m(t)$, the time-dependent $z$-component of 
the fiber center-of-mass position. Examples are shown in Fig.~\ref{fig3a}. %

To describe the tumbling, we first define the end-to-end vector as the difference of the positions of the centers of the last and first beads. Then, we determine the flipping instants $t_f$ as the moments when the end-to-end vector is oriented along $z$. At a flipping instant, the center-of-mass lateral position is denoted as $z_f\equiv z_m(t_f)$. 
The tumbling time $\tau$ associated with $t_f$ is defined as the difference between $t_f$ and the next consecutive flipping instant.

\begin{figure}[ht!]
\includegraphics[width=8.05cm]{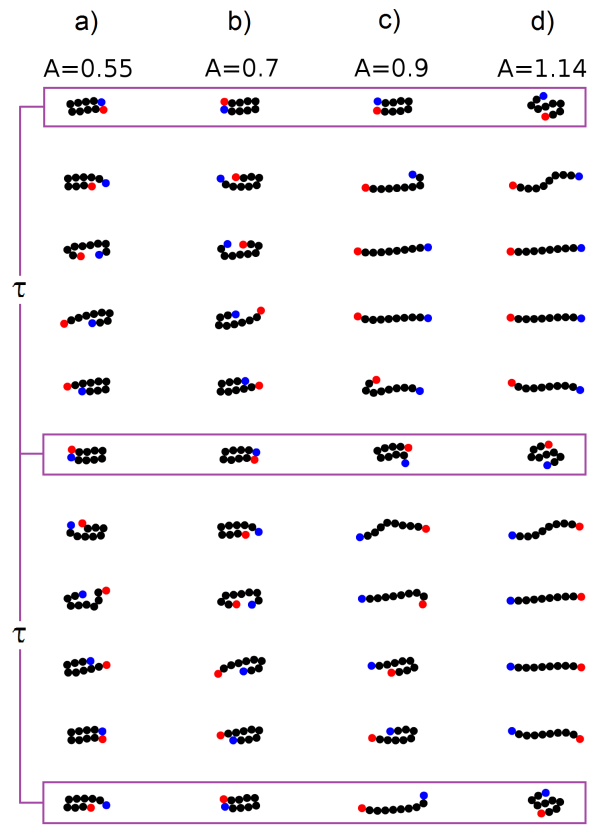}
\vspace{-0.1cm}
\caption{Shape evolution of fibers 
in shear flow for the fiber stiffness and tumbling time (a) $A\!=\!0.55$, $\tau\!=\!370$, (b) $A\!=\!0.7$, $\tau\!=\!352$, (c) 
$A\!=\!0.9$, $\tau\!=\!219$, (d) $A\!=\!1.14$, $\tau\!=\!206$. Snapshots are taken at equal time intervals.}
\label{fig2b}
\end{figure}

To analyze deformation of the fiber shape, we 
evaluate the time-dependent fiber curvature $\kappa(t)$ and fractional compression $\alpha(t)$, defined as  
\cite{Kantsler,Lindner,slowicka_lateral_2013}
\bee 
\kappa (t)\! &=& \!\frac{1}{N\!-\!2}\sum_{i=2}^{N-1} \frac{1}{r_i(t)},\hspace{0.4cm}{\alpha}(t)=1-\frac{\delta(t)}{(N\!-\!1)l_0},\;\;\; \;\;
\eee
where $r_i$ is the radius of the circle determined by the centers of three consecutive beads, and $\delta({t})$ is 
the fiber end-to-end distance  (i.e. the length of the end-to-end vector), initially equal to the equilibrium value 
 $(N-1)l_0$. 
Examples of $\kappa(t)$ and $\alpha(t)$, shown in Fig.~\ref{fig3a}, and the corresponding 
evolution of fiber shapes, illustrated at the snapshots from the simulations 
in Fig.~\ref{fig2b}, provide examples of 
different modes of the dynamics, which now will be introduced and discussed in details.

We first observe that fibers in Fig.~\ref{fig2b}(a,b) 
all the time stay coiled, but fibers in Fig.~\ref{fig2b}(c,d)
straighten out from time to time.  
For example, the fiber with $A\!=\!1.14$ 
straightens out along the flow, then forms a compact S-shape, straightens out again, and so on, as shown in Movie~(d). 
In contrast, the fiber with $A\!=\!0.7$ always stays coiled - it does not straighten out, as shown in Movie~(b). 
Its end-to-end vector rotates, with a well-defined tumbling time, but the trajectories of all the fiber beads stay close to each other, 
and the motion of the beads resembles  
tank treading observed for vesicles \cite{Misbah}.

The differences between the coiled and straightening out  modes 
are 
illustrated in Fig.~\ref{fig3a}. 
For the fibers which straighten out,  
the  curvature and fractional compression  
oscillate between a large value (corresponding to a compact shape) and almost zero (corresponding to almost linear 
 configuration). For fibers which do not straighten out and stay coiled, 
$\kappa(t)$ and $\alpha(t)$ also oscillate, but all the time their values remain large. 

To find the stiffness threshold for the coiled - straightening out transition, 
we 
perform simulations for a wide range of 
values of 
$A$, and  
analyze the minima $\kappa_{min}$ and 
$\alpha_{min}$ of the fiber curvature $\kappa(t)$ and  fractional compression $\alpha(t)$ with respect to 
the range of times 
\bee
t\in [4000, 16000],\label{tr}
\eee 
what corresponds to hundreds of tumbling times $\tau$. 
We start late enough to avoid initial transient effects, 
observed  
for certain values of $A$. The existence of such transients is straightforward for fibers which were initially 
aligned with the flow, but later remain coiled. 
\begin{figure}[b]
\includegraphics[width=8.4cm]{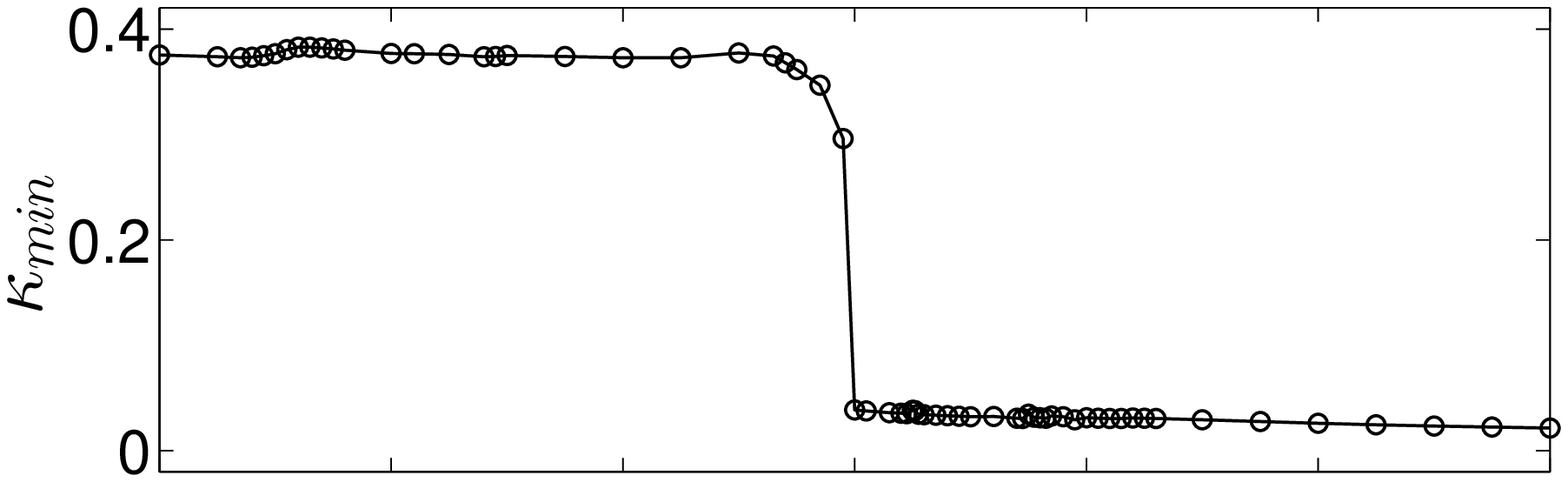}\;\;\;\;\\
\includegraphics[width=8.6cm]{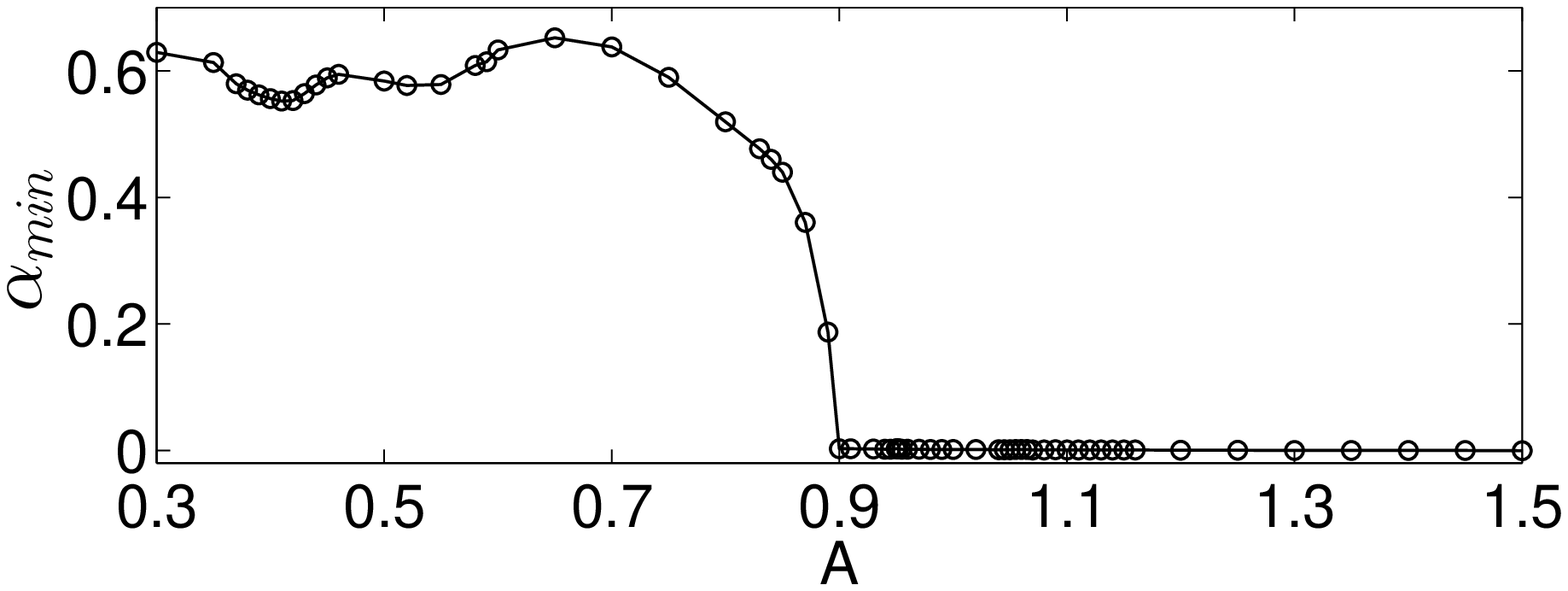}\;\;\;\;\\
\vspace{-0.2cm} 
\caption{Transition between the coiled and  straightening out modes:  
the minimal fiber curvature $\kappa_{min}$  and 
 fractional compression $\alpha_{min}$ 
versus $A$. 
}\label{st}
\end{figure}
The results are shown in Fig.~\ref{st}. 
A very sharp decrease 
of $\kappa_{min}$ and 
$\alpha_{min}$ at  $A_{CS} \!\approx \!0.9$ defines  
the stiffness threshold which separates the coiled mode of flexible fibers 
from the straightening out mode of stiff fibers.

Examples of the fiber dynamics, 
discussed above 
for $A\!=\!0.7$ and $A\!=\!1.14$, 
have a common property: they are strictly periodic. 
The fiber shape at time $t\!+\!\tau$ is the same as 
its shape at time $t$, rotated  by the angle $\pi$ with respect to $y$-axis. Owing to this symmetry, the consecutive tumbling times $\tau$ are equal to each other, 
 and there is no net migration across the flow.  
Indeed, as shown in Figs.~\ref{fig3a}$(b)$,$(d)$, a displacement $\Delta z_m$ of the fiber center-of-mass position $z_m$ at time $t$ is compensated by the opposite displacement $-\Delta z_m$ at time $t\!+\!\tau$. 

We will now analyze a different solution, called the migrating regular mode: superposition of periodic tumbling with a drift across the streamlines. 
An example  
with $A\!=\!0.55$ is shown in Figs.~\ref{fig3a}(a)-\ref{fig2b}(a) and Movie~(a).
For 
such a  dynamics, there is no rotational symmetry of shapes at time $t$ and $t+\tau$. The consecutive tumbling times $\tau_1$ and $\tau_2$ are not equal. 
The evolutions of shapes  
during 
$\tau_1$ and $\tau_2$ differ from each other, but 
during every second tumbling time are the same, with the period $\tau_1\!+\!\tau_2$. 
The asymmetry of the dynamics during two consecutive tumbling times results in a net fiber 
displacement along $z$, the same after each period, as shown  in Fig.~\ref{fig3a}$(a)$. 
\begin{figure}[b]
\includegraphics[width=8.6cm]{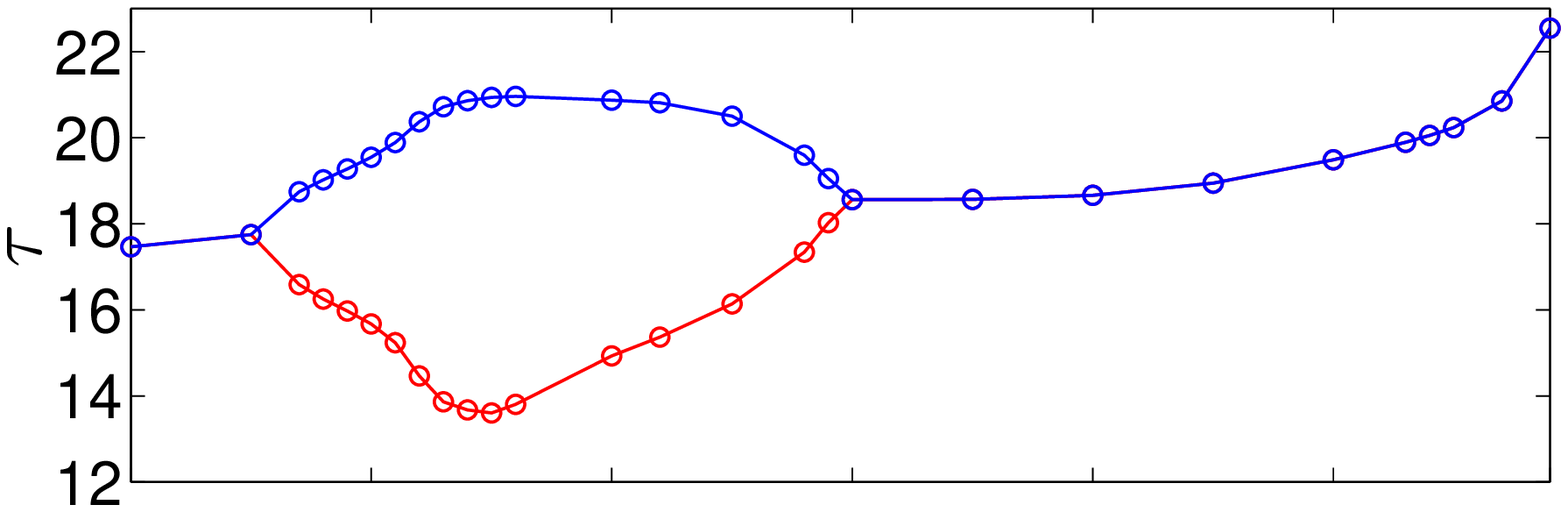}\\
\vspace{-0.47cm}
\includegraphics[width=8.6cm]{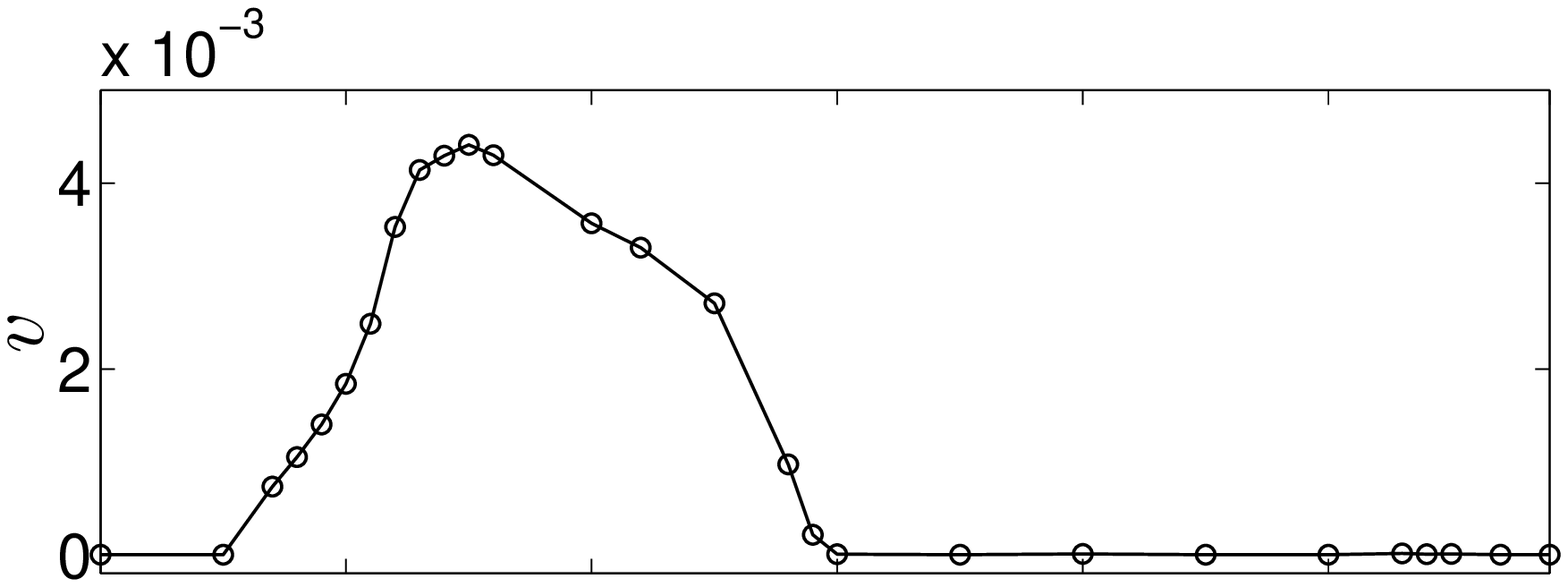}\\
\vspace{-0.43cm}
\includegraphics[width=8.6cm]{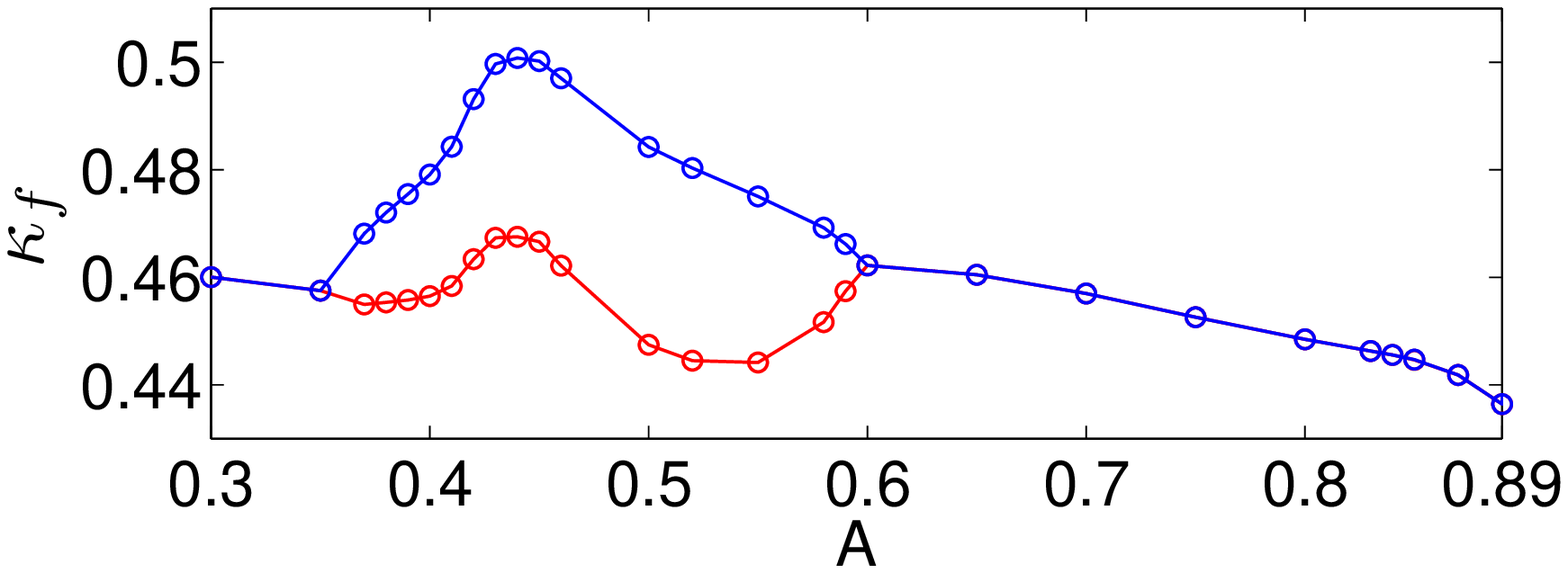}
\vspace{-0.8cm}
\caption{Period-doubling bifurcations 
between 
periodic 
and migrating regular 
modes: consecutive tumbling times $\tau$, migration velocity $v$ 
and  
curvature $\kappa_f$ 
versus~$A$.
}
\label{m-n}
\end{figure}

In Fig.~\ref{m-n}, we determine the stiffness thresholds for activation of the migrating 
regular modes.  
For each value of $A$ we plot all the tumbling times $\tau$ within the time range specified  in Eq.~\eqref{tr}, the mean migration velocity $v$  
(equal to the center-of-mass 
displacement $\Delta z_m$ across the flow   in the time  range specified in Eq.~\eqref{tr}, 
divided by the time difference), and the curvature $\kappa_f\equiv \kappa(t_f)$ at all the flipping instants $t_f$. 
At $A_{PDB}\!\!\approx\!0.35$ and 
$A_{PHB}\!\!\approx\!0.6$ we observe period-doubling and period-halving bifurcations \cite{bif} between the periodic mode (with a single value of the tumbling time and no migration) and the migrating regular mode (with two different
consecutive 
tumbling times $\tau_1$ and $\tau_2$). 

\begin{figure}
\includegraphics[width=8.6cm]{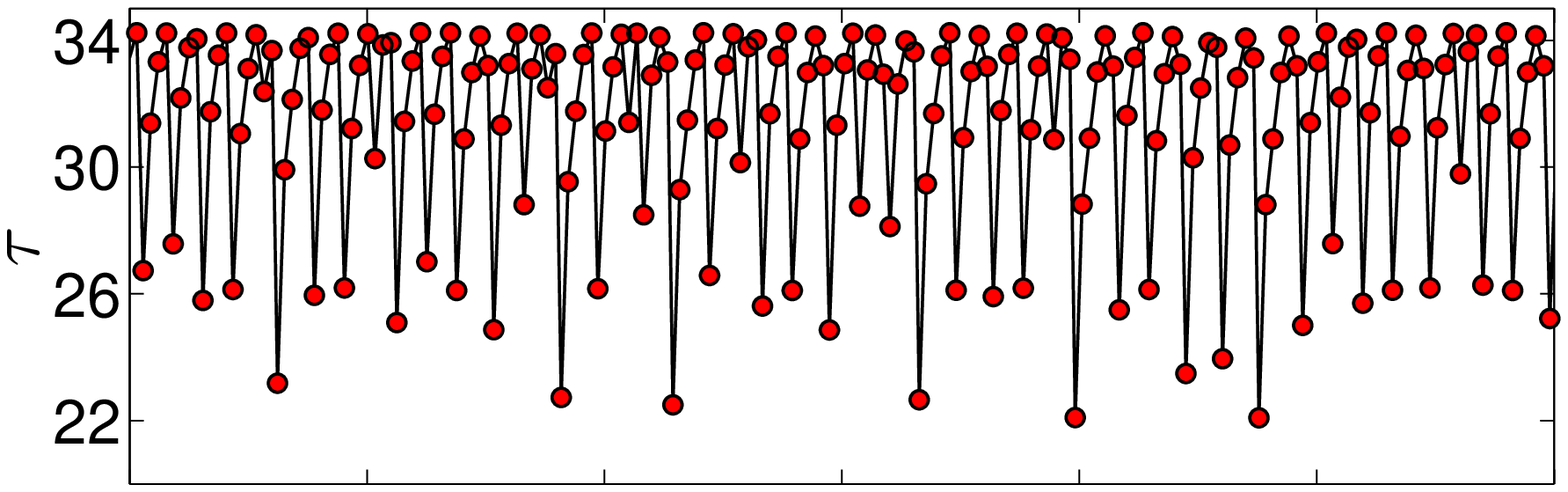} \\
\vspace{-0.29cm}
\includegraphics[width=8.6cm]{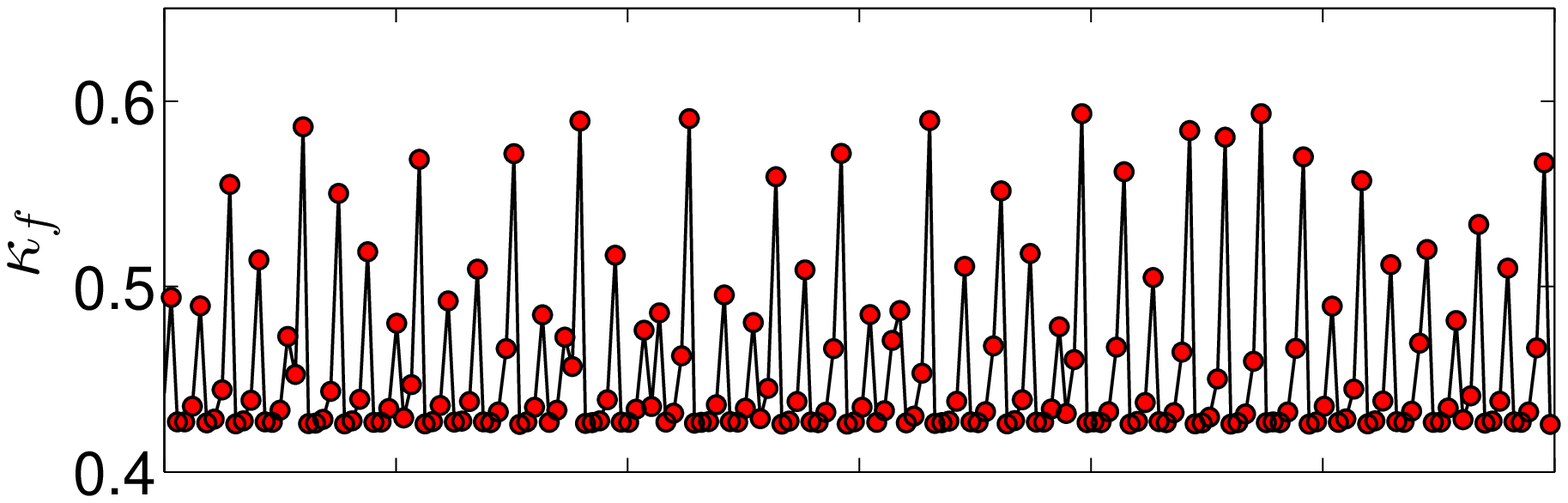} \\
\vspace{-0.24cm}
\includegraphics[width=8.6cm]{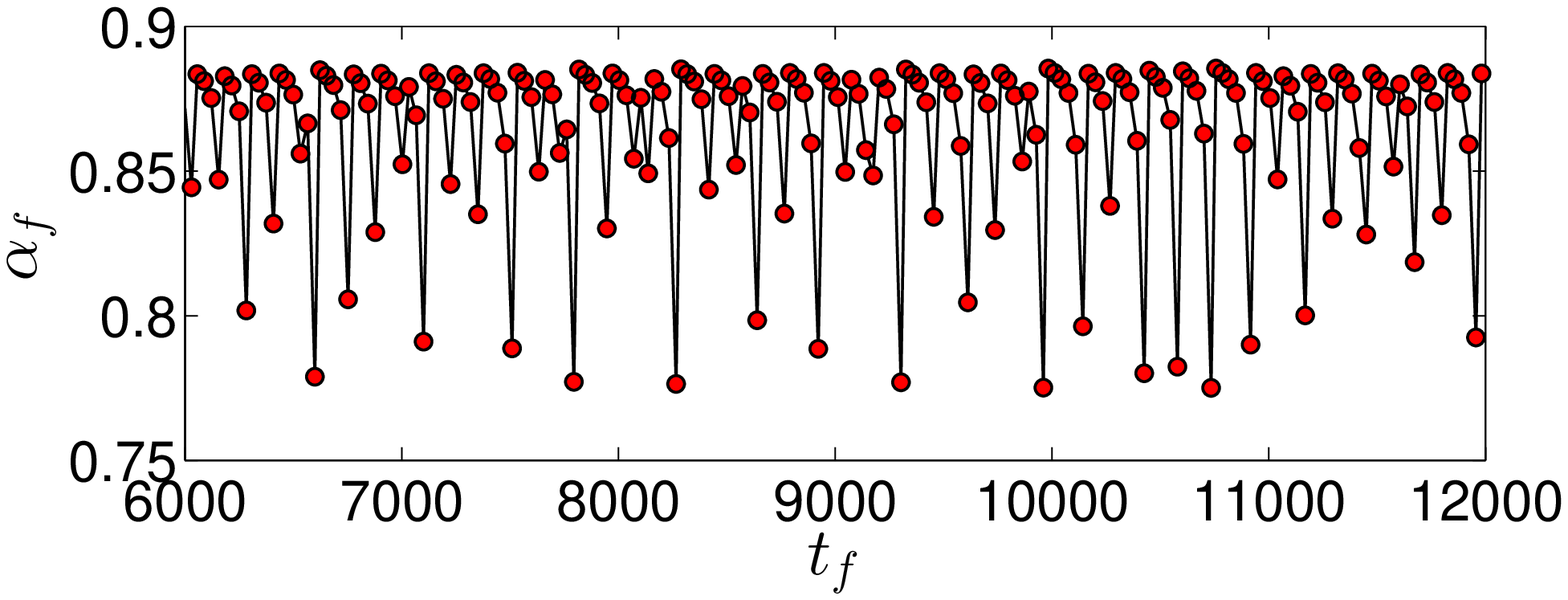} \\
\vspace{-0.3cm}
\caption{
Irregular mode 
($A\!=\!0.9$): consecutive tumbling times $\tau$, and fiber curvature $\kappa_f$ \& fractional compression $\alpha_f$ at consecutive flipping instants $t_f$.}\label{chao}
\end{figure}

\begin{figure}[b]
\includegraphics[width=8.6cm]{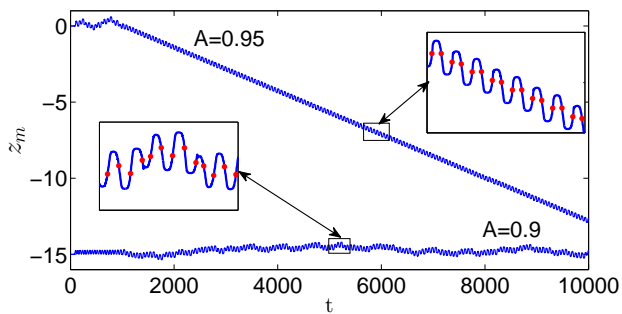}\\
\vspace{-0.3cm}
\caption{Examples of trajectories $z_m(t)$ in the chaotic range of the stiffness: 
irregular with $A\!=\!0.9$ (typical) and migrating regular with $A\!=\!0.95$ (exceptional).
}
\label{fig4}
\end{figure}

All the observed coiled solutions are periodic or migrating regular, as illustrated in Fig.~\ref{m-n}. However,  above the threshold $A_{CS}$ for the straightening out mode, a chaotic dynamics is observed. An example of  such an irregular solution with $A\!=\!0.9$ is shown 
 in Figs.~\ref{fig3a}$(c)$,~\ref{fig2b}$(c)$ and Movie~(c).  
A variability of the dynamics during different tumblings is visible. 
Subsequent tumbling times, and fiber curvature $\kappa_f$ \& fractional compression $\alpha_f$ at many consecutive flipping instants $t_f$ are shown in Fig.~\ref{chao}. 
Values of these three quantities evolve at random between certain upper and lower limits, what may correspond to one or more unstable migrating regular solutions.  The irregular, erratic  motion of the fiber center-of-mass 
is illustrated in Fig.~\ref{fig4}.


Chaotic trajectories have been observed for many values of the bending stiffness in the range $0.9 \!\le\! A\!\le\! 1.1$. Most of them are similar as for $A\!=\!0.9$, others contain transient parts with modulated oscillations (periodic-n solutions) or (occasionally) evolve towards migrating regular solutions, as illustrated in Fig.~\ref{fig4}. The evolution pattern is sensitive to small changes of the bending stiffness. 
In this range of $A$, solutions are also sensitive to the numerical accuracy, and therefore we do not perform any  systematic analysis. 

For $A\!\ge \!1.14$, solutions are periodic, with the basic properties discussed earlier for $A\!= \!1.14$ as an example. 

It is interesting to compare the tumbling times of flexible fibers with two reference solutions: rigid Jeffery's spheroids \cite{Jeffery} 
and rigid rods made of beads ($A\!=\!\infty$), both with the aspect ratios 
$r\!=\!10.09$: the same as our fiber straightened out at the equilibrium position. In our units, the reference tumbling time $\tau_s$ of the rigid spheroid \cite{Jeffery} is equal to $\tau_s \!=\! \pi(r\!+\!1/r)\!=\!32.0$. Using the {\sc hydromultipole} code with the truncation at the multipole order $L\!=\!10$, we evaluate 
the reference tumbling time $\tau_r \!=\!26.6$ of a rigid fiber made of beads. 

Coiled flexible fibers have an effective aspect ratio 
smaller than $r$. Therefore, it is reasonable that 
their tumbling times in Fig.~\ref{m-n} are below $\tau_r$. 
However, surprisingly, we have found that for flexible fibers which straighten out, the tumbling times can 
 exceed $\tau_r$, even though their mean curvature is large and the mean geometrical aspect ratio is smaller than $r$. 
In the chaotic range, the tumbling times fluctuate up to $\tau \!\approx \!35$, a value larger than $\tau_s\! >\! \tau_r$, 
and in the periodic range with $A\!\ge \!1.14$, the tumbling times $\tau\!\ge \!\tau_s \!>\!\tau_r$. In a sense, it takes an additional time to change shape while tumbling.

\begin{figure}[ht]
\includegraphics[width=8.5cm]{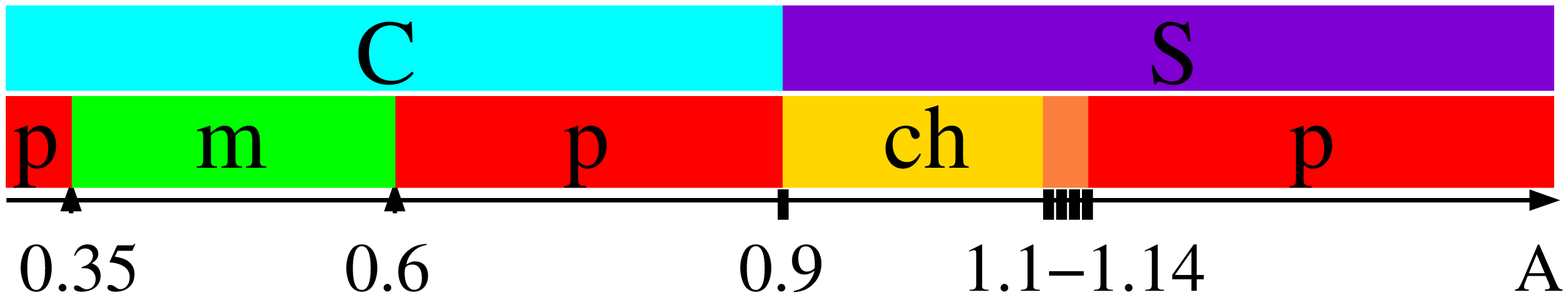}
 \caption{Dynamics of 
fibers 
with a different stiffness A. Modes: 
{\bf C} - coiled, 
{\bf S} -  
straightening out, {\bf p} - periodic, 
{\bf m} - migrating regular,
{\bf ch} - chaotic. Thresholds: 
$\blacktriangle$  - period-doubling bifurcation, 
$\bm{\mathsf{I}}\!\!\bm{\mathsf{I}}\!\bm{\mathsf{I}}\!\!\bm{\mathsf{I}}\!\!\bm{\mathsf{I}}\!\!\bm{\mathsf{I}}\!\!\bm{\mathsf{I}}\!\!\bm{\mathsf{I}}$  - transition to chaos. 
}\label{diagram}
\end{figure}

Concluding, the motion of 
flexible fibers with open ends and moderate aspect ratio in shear flow has been determined for a wide range of their bending stiffness $0.3 \le A \le 50$, assuming the initial alignment with the flow. 
Different modes of the dynamics have been observed for different values of $A$, as summarized in Fig.~\ref{diagram}. Period doubling bifurcations and transition to chaos have been found, in analogy to previous studies of different deformable objects: flexible knotted fibers \cite{Yeomans} and vesicles~\cite{Misbah_chaos}. 

\section*{Contributions}
A. M. S. performed all the computations for flexible fibers and showed their results in figures 1-6 and movies (a-d). E. W. provided the numerical code HYDROMULTIPOLE and the data for a rigid fiber. M. L. E.-J. designed and supervised the project, analyzed the results and wrote the paper. 

\acknowledgments
This work was supported in part by the Polish National Science Centre, 
Grant No. 2011/01/B/ST3/05691. M.L.E.-J. benefited from scientific activities of the COST Action MP1305. Computations were in part performed using the cluster ``Grafen'' (IPPT PAN, Biocentrum Ochota).  M.L.E.-J. thanks Marek Kocha\'nczyk for numerical tests and discussions. 

\end{document}